\documentclass[pre,aps,twocolumn,showpacs,floatfix,superscriptaddress]{revtex4-2}
\usepackage{graphicx}
\usepackage{bm}   
\usepackage{color}
\usepackage{amssymb}
\usepackage{amsmath}
\usepackage{ulem}
\usepackage[usenames,dvipsnames]{xcolor}
\usepackage{ragged2e}
\usepackage{adjustbox}
\usepackage{multirow}
\usepackage{booktabs} 
\usepackage{tcolorbox}
\usepackage{tabularx}
\usepackage{array}
\usepackage{colortbl}
\usepackage{hyperref}
\usepackage{float}
\hypersetup{colorlinks=true, citecolor=blue, urlcolor=blue, linkcolor=blue}
\tcbuselibrary{skins}

\newcolumntype{Y}{>{\raggedleft\arraybackslash}X}

\tcbset{tab1/.style={\centering, fonttitle=\bfseries\small,fontupper=\normalsize\sffamily,
		colback=yellow!10!white,colframe=red!75!black,colbacktitle=Salmon!40!white,
		coltitle=black,center title,freelance,frame code={
			\foreach \n in {north east,north west,south east,south west}
			{\path [fill=red!75!black] (interior.\n) circle (3mm); };},}}

\tcbset{tab2/.style={enhanced,fonttitle=\bfseries,fontupper=\small\sffamily,
		colback=yellow!10!white,colframe=red!50!black,colbacktitle=Salmon!40!white,
		coltitle=black,center title}}

\begin{document}

\title{Extreme mixed-mode oscillatory bursts in the Helmholtz-Duffing oscillator}
\author{S. Sudharsan}
\affiliation{Physics and Applied Mathematics Unit, Indian Statistical Institute, Kolkata 700108, India}
\author{Subramanian Ramakrishnan}
\affiliation{Department of Mechanical and Aerospace Engineering, University of Dayton, Dayton, OH 45469, USA}

\begin{abstract}
We report a new type of extreme event - \textit{extreme irregular mixed-mode oscillatory burst} - appearing in an asymmetric double-welled, driven Helmholtz-Duffing oscillator. The interplay of cubic and quadratic nonlinearities in the system, along with the external drive, contributes to this type of unusual extreme event. These extreme events are classified using the peak-over threshold method and found to be well-fitted with the generalized extreme value distribution. The asymmetry in the depth of one of the potential wells allows the oscillator to exhibit rare irregular mixed-mode oscillations between the two wells, manifesting as extremely irregular mixed-mode oscillatory bursts. Furthermore, we also find that such an irregular transition between the two potential wells occurs during the up phase of the periodic external drive. Importantly, we also find that the system velocity can be tracked and utilized as a reliable lead indicator of the occurrence of this novel type of extreme event. 
\end{abstract}
\maketitle

\section{Introduction}
Extreme events in dynamical systems \cite{farazmand2019review,sayantan2021review} are marked by the sudden onset of significant deviations from the expected mean values of system response characteristics \cite{dysthe2008hs}. Whilst occurring relatively infrequently, extreme events are yet recurrent in a broad class of natural \cite{albeverio2006extreme} and engineered systems \cite{Joo2017eng}, and their emergence often has significant consequences \cite{shenoy2022pnas}. Examples include natural events such as earthquakes \cite{qi2024seismic} and tsunamis \cite{ROBKE2017296}, biological events such as pandemics \cite{machado2020rare,singh2025epidemics,singh2024instabilities, MAJID2021322} and epileptic seizures \cite{frolov2019statistical}, and those in engineered systems such as power grids \cite{suncascading}, cyber and transportation networks \cite{schneider2011cyber,chen2015extreme,singh2022emergence}, and stock markets \cite{Sornette+2003}. The profoundly disruptive effects \cite{bell2018changes,devinsky2018epilepsy,doocy2013human} of extreme events on a diverse spectrum of dynamical systems \cite{sayantan2021review} underscore the importance of elucidating the underlying mechanisms \cite{farazmand2019review} that engender such events and developing reliable forecasting methods \cite{sayantan2021review} based on available information, such as historical time series of system response \cite{hallerberg2008prediction,bialonski2015prediction,meiyazhagan2021model}. Understanding and predicting extreme events in dynamical systems has therefore been described as a grand challenge \cite{majda2019}, and the topic has attracted much recent research that includes analytical \cite{lucarini2016extremes,suresh2018timedelay}, experimental \cite{premraj2021dragon}, and data-driven approaches \cite{bialonski2015prediction,meiyazhagan2021prediction,meiyazhagan2021prediction,ray2021prediction, SHARMA2018139}.

In this article, we investigate extreme events in the Hemholtz-Duffing (H-D) oscillator, a nonlinear dynamical system of interest both from a theoretical perspective and in applications. The analysis of instabilities illuminates the characterization of extreme events in nonlinear oscillators (see, for instance, the review article \cite{sayantan2021review}), and we adopt this approach in our analysis. Extreme events occur when the system trajectory traverses instability regions in the phase space. Specifically, instabilities push the trajectory far away from attractor regions, facilitating rare and long excursions that result in extreme events. However, only a few types of instabilities have thus far been reported as potential sources for producing such long excursions. In this context, Kingston et al. \cite{kingston2017extreme} reported extreme events under the influence of external forcing in a nonlinear system of the Li\'enard type, introduced by Chandrasekar et al. \cite{chandrasekar2005}. In addition, interior crisis (IC) and the Pommeau-Maneville intermittency (PMI)\cite{kingston2017extreme,kingston2020extremelieneard} were identified as specific mechanisms that support the onset of extreme events in this system. We note that an IC is associated with the sudden expansion of a chaotic attractor from a low-amplitude chaotic attractor. At the same time, a PMI corresponds to a similar expansion from a single periodic orbit. Interestingly, both IC and PMI were subsequently found to induce extreme events in a broad class of chaotic systems such as the Ikeda map \cite{ray2019extreme}, a non-polynomial mechanical system \cite{sudharsan2021symmetrical}, the Bonhoeffer–van der Pol oscillator \cite{bhagyaraj2023super,thangavel2021extreme,wei2023mmoextreme}, the El Ni\~no southern oscillation model \cite{ray2020understanding}, a population model \cite{vijay2024extreme}, a SQUID trimer \cite{vijay2024squid}, the Sprott-Like Model \cite{thamilmaran2024sprott}, a nonhyperbolic chaotic
system \cite{vijay2024extreme}, and a memristor-based Hindmarsh–Rose neuron model \cite{vijay2024transition,vijay2023superextreme}. The mechanisms above have also been found to induce extreme events in stochastic systems \cite{Franzke2017noise, Zhao2023noise}. Moreover, variations in the mechanism of sudden expansion or destruction of chaotic attractors have also been identified. These include hyperchaotic expansion \cite{kingston2023hyperchaos,kingston2023hyperchaos1}, gradual changes or alternate expansion and contraction of a chaotic attractor as a result of changes in the velocity \cite{sudharsan2021emergence}, and extreme events reported in a microelectromechanical (MEMS) system and a $CO_2$-based laser model \cite{suresh2018mems} arising from trajectories in the vicinity of discontinuous boundaries. While the instabilities involved in all of the cases above are associated with chaotic strange attractors, recent work reported long excursions of trajectories in two distinct types of nonlinear oscillators with non-chaotic strange attractors \cite{kaviya2022extreme,prem2023quasi}. Finally, in addition to isolated systems \cite{sayantan2021review}, extreme events have also been reported in coupled and complex networks \cite{ray2022networks,ansmann2013extreme,ray2020josephson,sree2024extreme,shashangan2024mitigation,roy2023extreme,roy2024extreme} as the body of literature on the topic continues to grow at a tremendous rate.

In contrast to the above cases, where extreme events manifest as rare, large-amplitude spikes in an otherwise small-amplitude time series, here we report a novel type of extreme event, which we term \textit{extreme irregular mixed mode oscillatory burst}. We find this in a driven H-D oscillator, a system characterized by the presence of both quadratic and cubic nonlinear terms in the potential energy and is also utilized in several applications \cite{rega1995bifurcation}. Akin to the Duffing oscillator, geometrically, the potential energy curve of an H-D oscillator can have either a double well (or hump) or a single well (hump), depending on the choice of the parameters. However, a key difference is the asymmetry in the shape of the potential well depending on the interplay between the two types of nonlinearity, especially in the double well case. We find that this asymmetry plays an important role in generating extreme bursts in the H-D oscillator. In particular, when the oscillator is driven by external forcing, we find that for a specific set of parameter values (corresponding to an asymmetric double well), the system exhibits bursts of \textit{irregular mixed-mode oscillations} (MMO) representing extreme events. This rare burst of irregular MMO occurs when the trajectories oscillate irregularly back and forth between the two wells. In other words, the system exhibits periodic MMO when the dynamics are restricted within a single, deeper well (corresponding to higher values of the quadratic nonlinearity coefficient). However, for lower values of the quadratic nonlinearity coefficient, the system starts exhibiting irregular MMO bursts, indicating excursions to the other shallower well. This intermittent, irregular motion between the wells corresponds to extreme events.

To further highlight the novelty of the mechanism that generates the extreme events reported in this article, we now discuss bursting oscillations that have been recently reported in certain mechanical systems. In Ref.~\cite{kaviya2022extreme}, the authors show that a parametrically driven Rayleigh-Li\'enard system exhibits extreme bursting oscillations (due to the pulse-shaped explosion of the fixed points). The trajectories evolve chaotically with both small amplitude and rare large amplitude spikes; the large amplitude spikes, occasionally, correspond to extreme events. Moreover, in \cite{wei2023mmoextreme}, the authors report extreme events in a slow-fast system where, as a result of interior crisis, amidst typical MMO, isolated large spikes rarely emerge as extreme events. Additionally, in Ref.~\cite{kaviya2024extreme}, asymmetry in a double well potential is reported to produce extreme events in a system of the Li\'enard type; however, they occur as rare chaotic large amplitude spikes in otherwise small amplitude chaotic dynamics. In other words, the trajectories oscillate inside the left well chaotically and rarely visit the other well. It bears reiteration that in all the aforementioned cases, extreme events occur as large amplitude spikes in chaotic time series comprising both small and large amplitude spikes. However, in the case of the H-D oscillator, we find extreme events that emerge as sudden bursts of irregular mixed-mode oscillations in the midst of periodic mixed-mode oscillations. Structurally, the shape of the extreme events also stands distinct from all the other reported cases. Moreover, the causative mechanism behind the extreme MMO burst in the H-D oscillator that we identify and present in this article has not been reported in the literature. In fact, to the best of our knowledge, extreme irregular mixed-mode oscillatory bursts themselves have yet to be investigated from the standpoint of extreme events. These aspects constitute the key novel contributions of this article.

The H-D oscillator arises in a broad variety of contexts, including nonlinear structural dynamics applications \cite{rega1995bifurcation}, studies of optomechanical systems, energy harvesting, pull-down instability, breaking symmetry, functionally-graded systems, composite plates, and snap-through mechanisms. In addition, several phenomena in plasma physics are modeled using the Gardner equation and the extended KdV-Burger equation, which, under appropriate transformations, respectively yield the undamped and damped H-D oscillator equations \cite{el2021novel}. Therefore, studies of extreme events in the H-D oscillator are also of interest from the viewpoint of applications.

The rest of the article is set as follows. In Sec.~\ref{model}, we introduce the H-D model, numerically solve for the dynamics, establish the emergence of extreme events in the system, and characterize the events using appropriate statistical measures. Next, in Sec.~\ref{mechanism}, we elucidate the mechanism that engenders extreme events in this system. Focusing next on predicting the emergence of the events, in Sec~\ref{ri_sec}, we present results that highlight velocity as a reliable lead indicator in this context. The article concludes in Sec.~\ref{discuss} with a summary of the results and an outlook for further research.

\section{Model and Dynamics}
\label{model}
We consider the Helmholtz-Duffing oscillator with the potential energy $V(x)=c_1\frac{x^2}{2}+c_2\frac{x^3}{3}+c_3\frac{x^4}{4}$. The equation of motion is:

\begin{eqnarray}
	\ddot{x} + \mu\dot{x} + c_1 x + c_2 x^2 +c_3 x^3 = F \mathrm{cos} \omega t,
    \label{mastereqn}
\end{eqnarray}

where $\mu$ is the damping coefficient, $c_1$ the linear stiffness coefficient, and $c_2$ and $c_3$ are the nonlinear stiffness coefficients. $F$ and $\omega$ are the amplitude and frequency, respectively, of external excitation. For the analysis of extreme events in Eq.~(\ref{mastereqn}), we fix the values of these parameters as $\mu=0.1$, $c_1=-1.0$, $c_3=534.53$, $F=0.5$, and $\omega=0.42$. The signs of $c_1$, $c_2$ and $c_3$ are chosen appropriately to yield an asymmetric double well potential. The asymmetry generated by the variation in depth of one of the potential wells is calibrated by the value of $c_2$. Therefore, we take it as the bifurcation parameter in the present work. The numerical integration of Eq.~(\ref{mastereqn}) is carried out with the initial conditions $x(0)=0.01,\dot{x}=0.01$ using the fifth-order Runge-Kutta Fehlberg algorithm (RKF45) with a fixed step size of 0.01. For the numerical integration, we convert Eq.~(\ref{mastereqn}) to a system of first-order differential equations as

\begin{eqnarray}
    \dot{x} &=& y \nonumber \\
    \dot{y} &=& F \mathrm{cos} \omega t - \mu y - c_1x - c_2x^2 - c_3x^3 
\label{spliteqn}
\end{eqnarray}

Here $y$ denotes the velocity. After numerically integrating Eq.~\ref{spliteqn}, we classify the oscillator dynamics as extreme or otherwise using the Peak Over Threshold (POT) criterion \cite{sayantan2021review}. The POT method involves determining the qualifier threshold ($x_{th}$) for extreme events using the following relation.

\begin{eqnarray}
    x_{th} = \langle x_{peaks} \rangle + n\sigma, \quad n \in \mathbb{R}~~\backslash~~\{0\}~~\text{and}~~n>1. \nonumber \\
    \label{threshold}
\end{eqnarray}

Here $\langle x_{peaks} \rangle$ denotes the average value of the local maximum, and $\sigma$ the corresponding standard deviation. The real number $n$ can take any value except $0$ and $\pm 1$. The choice of $n$ is arbitrary. Typically, a good choice would be $n\geq4$, marking a deviation from the central tendency adequate to capture rare events. Throughout the present work, we fix $n=5$. In the subsequent parts of this article, we first classify the dynamics of extreme oscillations, statistically characterize them, and determine the mechanism behind their emergence.

\begin{figure*}[!]	\includegraphics[width=1.0\linewidth]{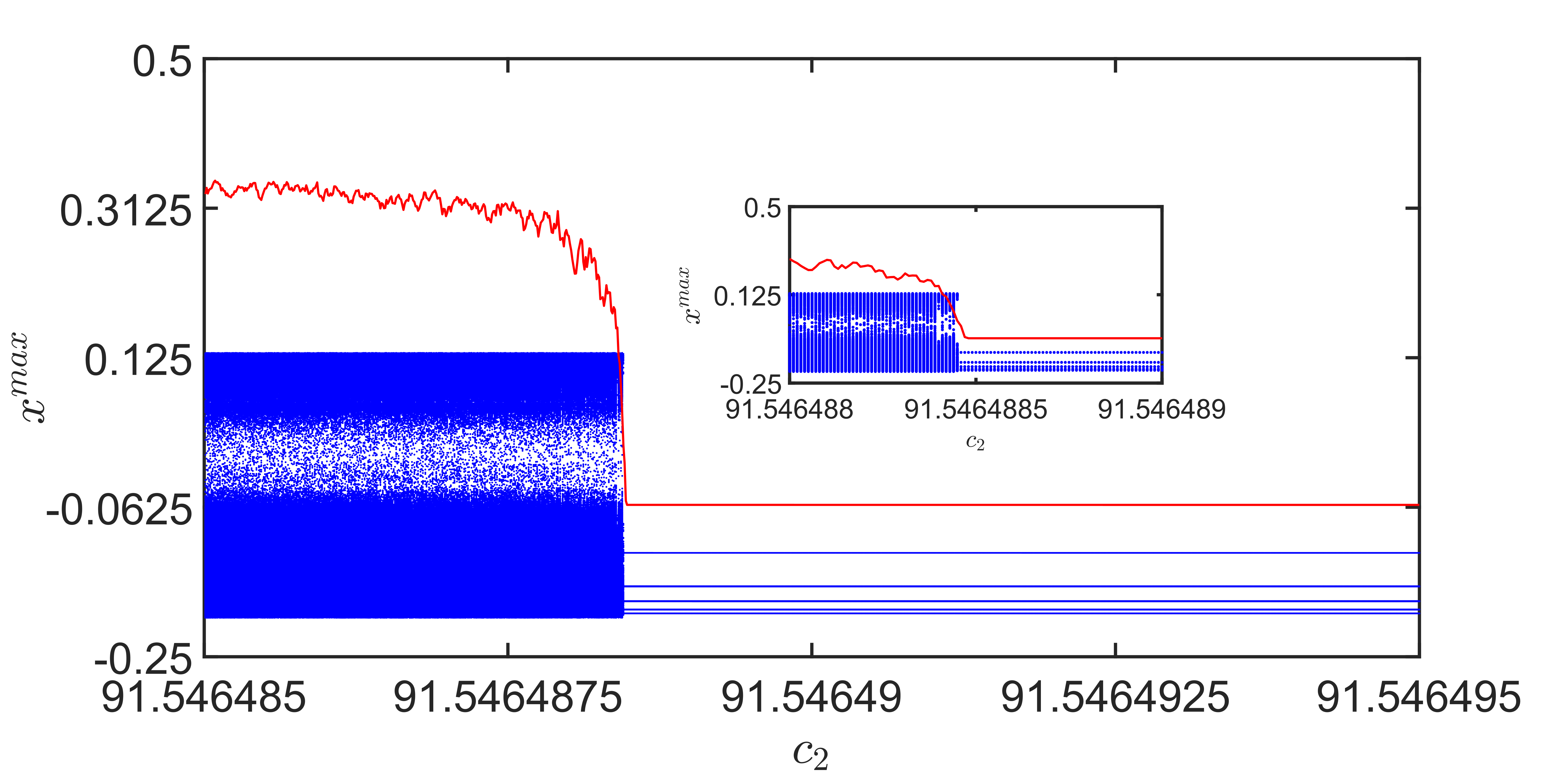}
\caption{Bifurcation plot (blue points) of Eq.~\ref{mastereqn}, with respect to $c_2$. Red solid line represents the $x_{th}$ with $n=5$. The inset figure provides better visualization of the peaks crossing the threshold for finer values of $c_2$.}
\label{figbifur}
\end{figure*}

To visualize the change in dynamics and to clearly identify the emergence of extreme events in the system (\ref{mastereqn}) with respect to $c_2$, we plot in Fig.~\ref{figbifur}, the corresponding bifurcation diagram of the system (\ref{mastereqn}) by collecting the local maximum values. From Fig.~\ref{figbifur}, we observe that when the value of $c_2$ is monotonically increased, at $c_2=91.5464885$, the oscillator undergoes a bifurcation from exhibiting chaotic dynamics to multiperiodic oscillations. Specifically, just before the transition, we note that for a few values of $c_2$ (to the left side of the transition), the threshold value (red solid line) cuts the bifurcation diagram in a way that clearly illustrates the presence of peaks above the threshold value. These peaks above the threshold are extreme events. For clearer visualization, an inset figure showcasing the same for finer values of $c_2$ is also presented.

\begin{figure*}[!]
\includegraphics[width=1.0\linewidth]{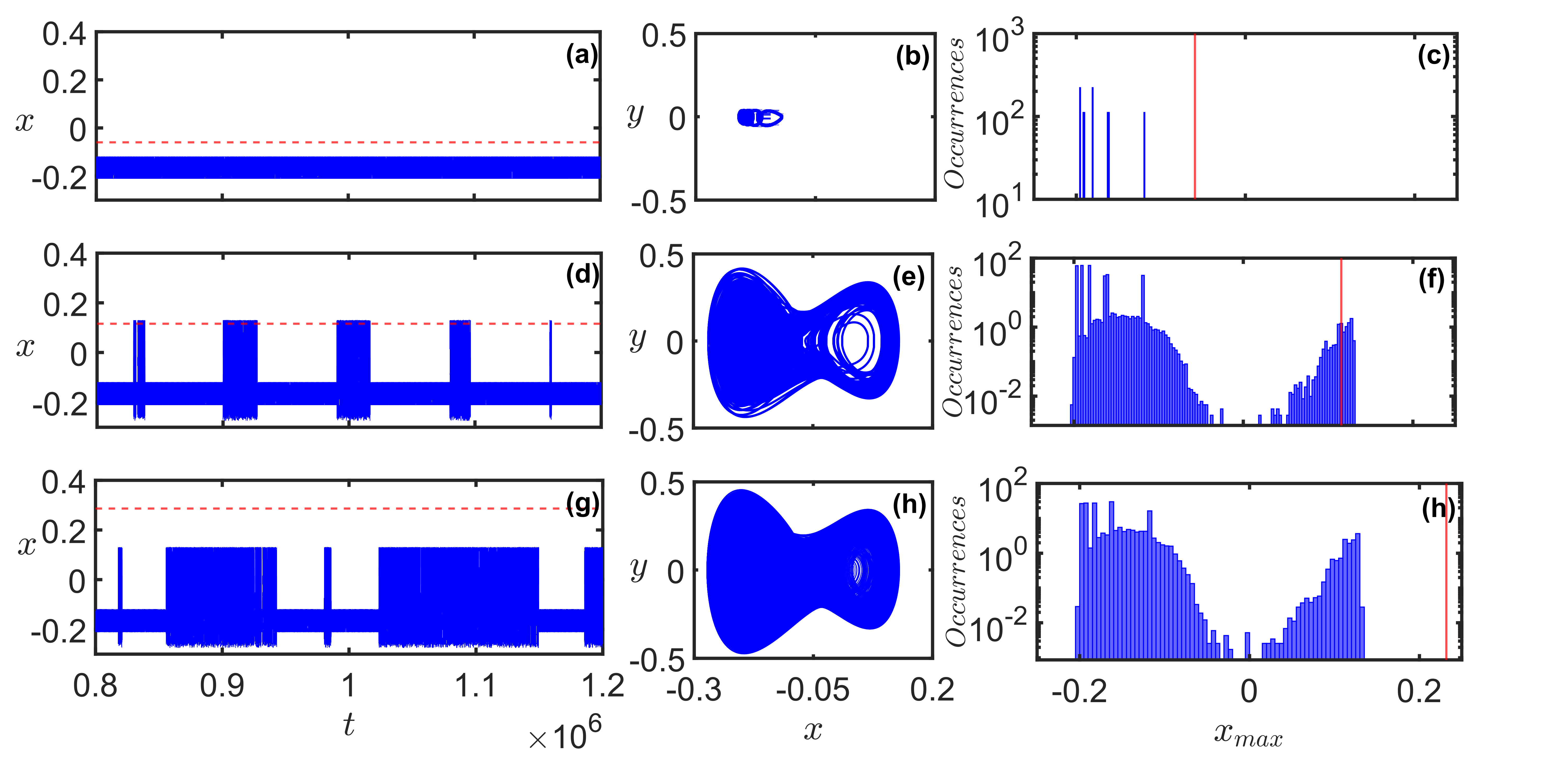}
\caption{Time series (Column-1), Phase portraits (Column-2), and histogram of peaks (Column-3) of system (\ref{mastereqn}) for $c_2=91.5464885$ in subplots (a-c), for $c_2=91.5464884$ in subplots (d-f), and for $c_2=91.5464883$ in subplots (g-h). Horizontal red dotted line in Column-1 and the vertical red solid line in Column-3 are the $x_{th}$.}
\label{tscollage}
\end{figure*}

We now take a closer look at the dynamics before, during, and after the transition, using time series, phase portraits, and histograms of events in Fig.~\ref{tscollage}. In the figure, the horizontal red dotted line in Column-1 and the vertical red solid line in Column-3 correspond to the threshold value calculated using Eq.~(\ref{threshold}). Corresponding to the value of  $c_2=91.5464885$ the oscillator exhibits multiperiodic oscillations, as evident from Row-1 of Fig.~\ref{tscollage} (subplots (a-c)). In particular, the higher resolution segment provided in Fig.~\ref{tscollage}(a) reveals a densely packed time series corresponding to multiperiodic oscillations. Moreover, these oscillations are also observed in the phase portraits in Fig.~\ref{tscollage}(b). In addition, the distinct straight lines in Fig.~\ref{tscollage}(c) also clearly indicate the peaks of the multiperiodic dynamics. Upon decreasing the value of $c_2=91.5464884$, (the transition point in the Fig.~\ref{figbifur}), one further observes from the time series in Fig.~(\ref{tscollage}(d)) a sudden burst of higher amplitude oscillations that are interspersed between lower amplitude oscillations. This conspicuous burst lasts only for a short period of time (in a pulsed manner). A remarkable feature here is that whenever the oscillations burst to higher amplitude, they all cross the threshold value. This can also be visualized in a different manner in the phase portrait in Fig.~\ref{tscollage}(e), where a double scroll attractor with a dense left and a sparse right scroll emerges. Here, the left scroll represents oscillations with lower amplitude, and the right scroll represents those with higher amplitudes. Furthermore, in the histogram of extreme events (Fig.~\ref{tscollage}(f)), this is reflected in the presence of peaks beyond the threshold. When we decrease the value of $c_2$ to $91.5464883$, we can observe, from the time series in Fig.~\ref{tscollage}(g), that the time taken by the trajectory for exhibiting burst of high amplitude oscillations is long-lasting and frequent. Together, these two features render the burst of high amplitude oscillations no longer rare. Therefore, the threshold value is well above the time series, and no trajectories can be found to cross the threshold; hence, no extreme events occur. Additional verification is also provided by the attractor in the phase portrait where the previously scarce right scroll has now become dense (c.f. Figs.~\ref{tscollage}(e)~$\&$~\ref{tscollage}(h)). Furthermore, in the histogram of peaks in Fig.~\ref{tscollage}(h), we find no peaks beyond the threshold. Increasing the value of $c_2$ increases the amount of time it takes for the oscillator to exhibit a high amplitude oscillation.

\begin{figure}[!h]
\includegraphics[width=1.0\linewidth]{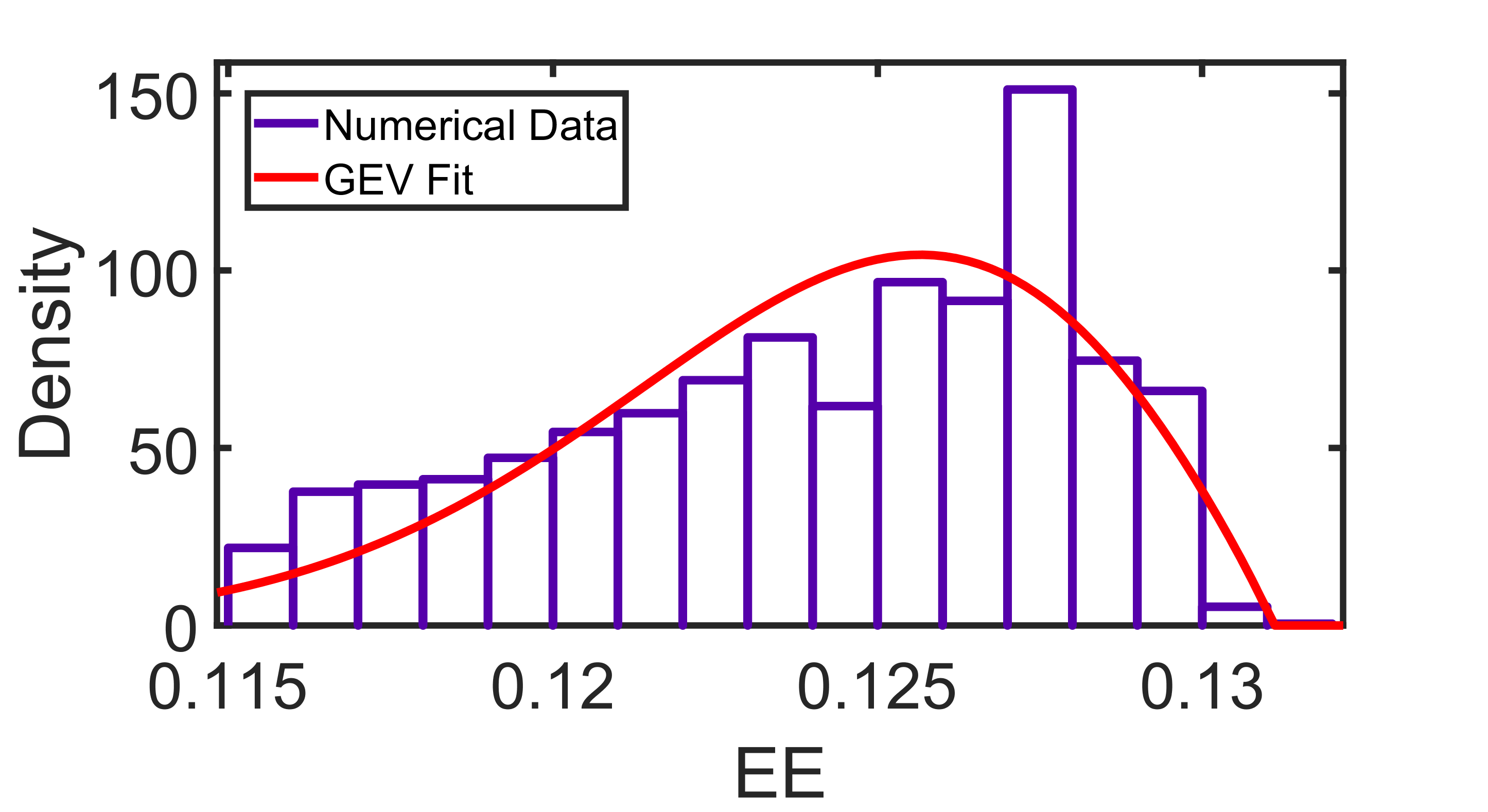}
\caption{Blocks represent the numerically generated extreme events data split into bins at $c_2=91.564884$. The red curve is the theoretical GEV fit (Eq.~\ref{eqnfit}) upon the numerically data.}
\label{fit}
\end{figure}

Notably, while all extreme events are rare events, the converse need not be true \cite{sayantan2021review}. Therefore it is important to confirm that the rare, high-amplitude burst oscillations that we uncovered in the H-D oscillator are indeed extreme events. To confirm this, we validate the events crossing the threshold using the generalized extreme value (GEV) distribution provided in Fig.~\ref{fit}.
The GEV distribution is mathematically represented by 
\begin{eqnarray}
	G(x) &=& \dfrac{1}{\beta}\exp\Bigg(-\Big(1+\gamma\dfrac{x-\alpha}{\beta}\Big)^{-\frac{1}{\gamma}}\Bigg) \nonumber \\&& \times \Big(1+\gamma\dfrac{x-\alpha}{\beta}\Big)^{-\frac{1}{\gamma}-1},
    \label{eqnfit}
\end{eqnarray}
for $\beta\neq0$ and $1+\gamma\dfrac{x-\alpha}{\beta} > 0$. Here, $\alpha\in \mathbb{R}$, $\beta >0$, and $\gamma\neq0$ are, respectively, the location, scale, and shape parameters. We note that, depending on the value of $\gamma$ (i.e. positive ($\gamma>0$) or negative ($\gamma<0$)), the distribution type also varies correspondingly as the Fr\'echet and Weibull distributions. Moreover, for $\gamma=0$, one obtains the Gumbel distribution. The values of the different parameters and the corresponding standard error of the distribution for extreme events appearing in the system (\ref{mastereqn}) are presented in Table.~\ref{table}. We have performed the GEV fit of the numerical data and estimated the values of the parameters using the MATLAB Distribution Fitter App. The bin sizes are fixed using the Freedman-Diaconis rule. From our estimates (Table.~\ref{table}), we find that the shape parameter is negative, which determines the fit to be a Weibull distribution.
\\
\begin{table}[!]
    \centering
    \begin{tabular}{|c|c|c|}
        \hline
       \textbf{Parameter} & \textbf{Estimate} & \textbf{Standard Error} \\
        \hline
        Shape ($\gamma$)   & -0.521961   & 0.00679248   \\
        \hline
        Scale ($\beta$)   & 0.00417023  & 5.36664e-05   \\
        \hline
        Location ($\alpha$)   & 0.123104 & 6.98099e-05   \\
        \hline
    \end{tabular}
    \caption{The estimate and the standard error obtained for the GEV distribution (Eq.~\ref{eqnfit}). Negative value of the shape parameter $\gamma$ indicates that the distribution is Weibull (type III). }
    \label{table}
\end{table}

\section{Advent of extreme events: The Mechanism}
\label{mechanism}
The most important questions that immediately follow the identification of extreme events in a dynamical system are related to understanding the dynamical origins of the events and the corresponding underlying mechanisms. For, in the absence of such understanding, extreme events only appear to be arbitrary, rare events. Accordingly, in this section, we investigate in greater detail the extreme event dynamics of the Helmholtz-Duffing oscillator (Eq.~(\ref{mastereqn})) and present the dynamical rationale that explains the emergence of extreme events in the oscillator.
\par
\begin{figure*}[!]
\includegraphics[width=1.0\linewidth]{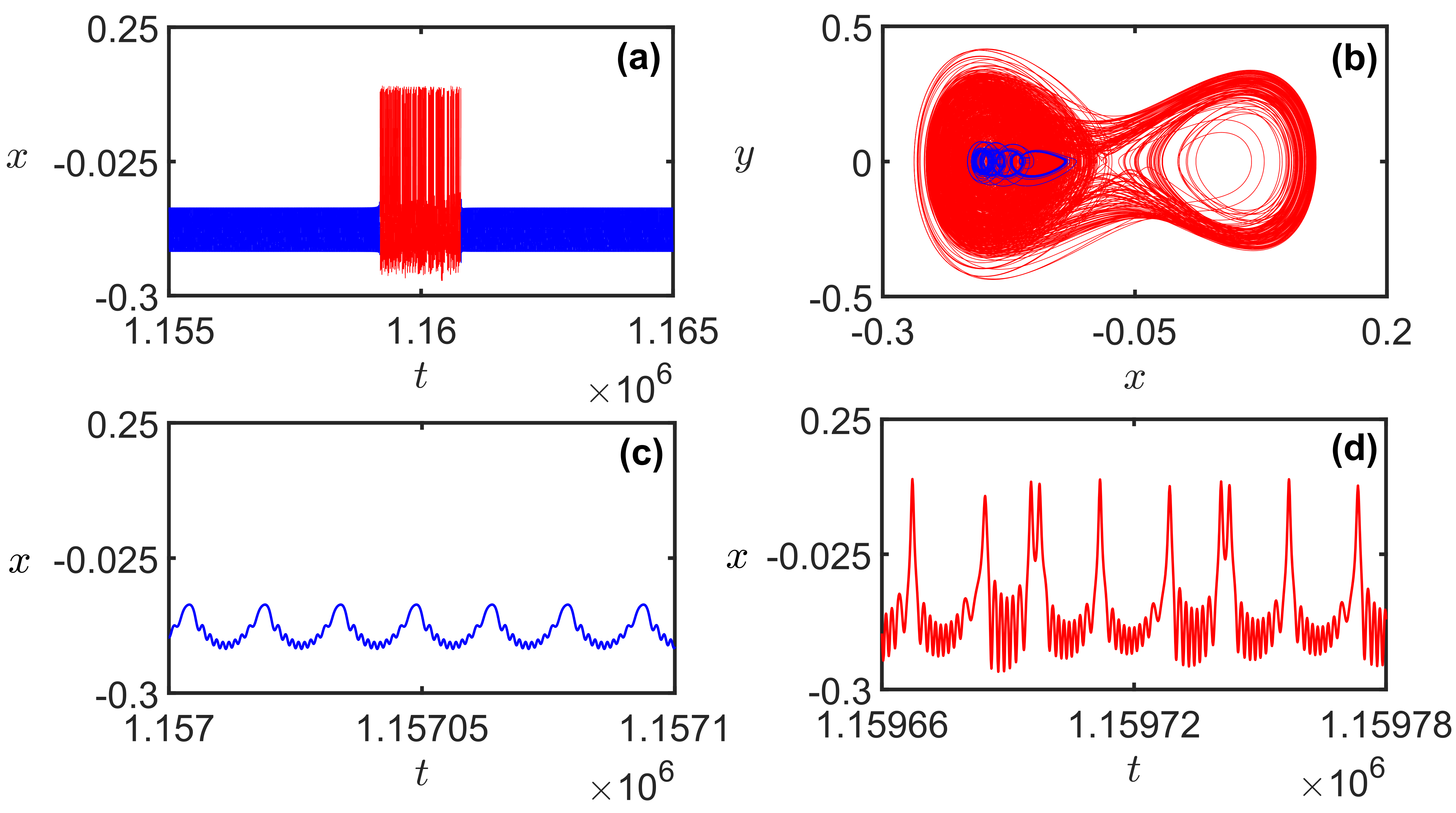}
\caption{(a) Times series in $x$ and (b) phase portraits distinctly depicting the non-extreme trajectories in blue and the extreme bursts in red. Individual time series: (c) non-extreme periodic MMO and (d) extreme burst of irregular MMO. All these represent the dynamics of system (\ref{mastereqn}) at $c_2=5.6494884$}
\label{fig-mecha}
\end{figure*}

 In Fig.~\ref{fig-mecha}, we plot both the system response time series and the phase portrait at $c_2=91.5464884$, the value of $c_2$ at which we observe extreme events. Note that the dynamics are chaotic for this value of $c_2$, as also confirmed by the value of the largest Lyapunov exponent, which is $0.0396177$. Recalling our observation that large amplitude oscillation bursts mark the emergence of extreme events, we begin our analysis by dissecting the time series separately for small and large amplitude oscillations. Accordingly, for better clarity and distinction, we plot in Fig.~\ref{fig-mecha}(a) the time series of the small amplitude and large amplitude oscillations in blue and red curves, respectively. We then observe that the trajectories represented by blue curves indicate bounded and regular motion, while the trajectories represented by red curves correspond to irregular motion. Moreover, we present the phase portraits of the small and large amplitude oscillations in blue and red colors, respectively, in Fig.~\ref{fig-mecha}(b). Note that the boundedness of the trajectories represented by the blue curves is quite apparent in this figure. Indeed, the boundedness of the blue curves reflects the system trajectories being confined to the interior region of the left potential well. In sharp contrast, the trajectories represented by red curves each have two scrolls (similar to the case of the Duffing oscillator); this clearly indicates that a trajectory represented by a red curve - no longer restricted to the single well - exhibits excursions to the right potential well. The cardinal point here is that such excursions of (red-colored) trajectories to the right potential well are rare enough; the rarity is not only established by the POT qualifier criterion but also asserted by the fact that the corresponding distribution is a Weibull distribution. Moreover, we note that for lower values of $c_2$, this transition between the two wells occurs too often and, therefore, does not qualify as an extreme event.
\par Interestingly, we also find that both small and large amplitude oscillations are, in fact, mixed-mode oscillations (MMO). In the case of small amplitude oscillations, regular MMO of value $1^8$ ($L^s$; $L$ and $s$ representing the number of large and small peaks, respectively) occur (Fig.~\ref{fig-mecha}(c)) whereas, in the case of large amplitude oscillations, irregular MMO are observed (see Fig.~\ref{fig-mecha}(d)). In other words, as $c_2$ is varied and the dynamics transition from multiperiodic oscillations to chaotic, the system trajectories also change from being confined to a single well to inter-well excursions. To summarize this aspect of the mechanism, variation in the quadratic nonlinear coefficient $c_2$ introduces an asymmetry in the Helmholtz-Duffing oscillator potential, which, in turn, induces transitions from periodic MMO to intermittent irregular MMO. These intermittent irregular MMO bursts are indeed the extreme events uncovered in our analysis.

\par
\begin{figure}[!h]
\includegraphics[width=1.0\linewidth]{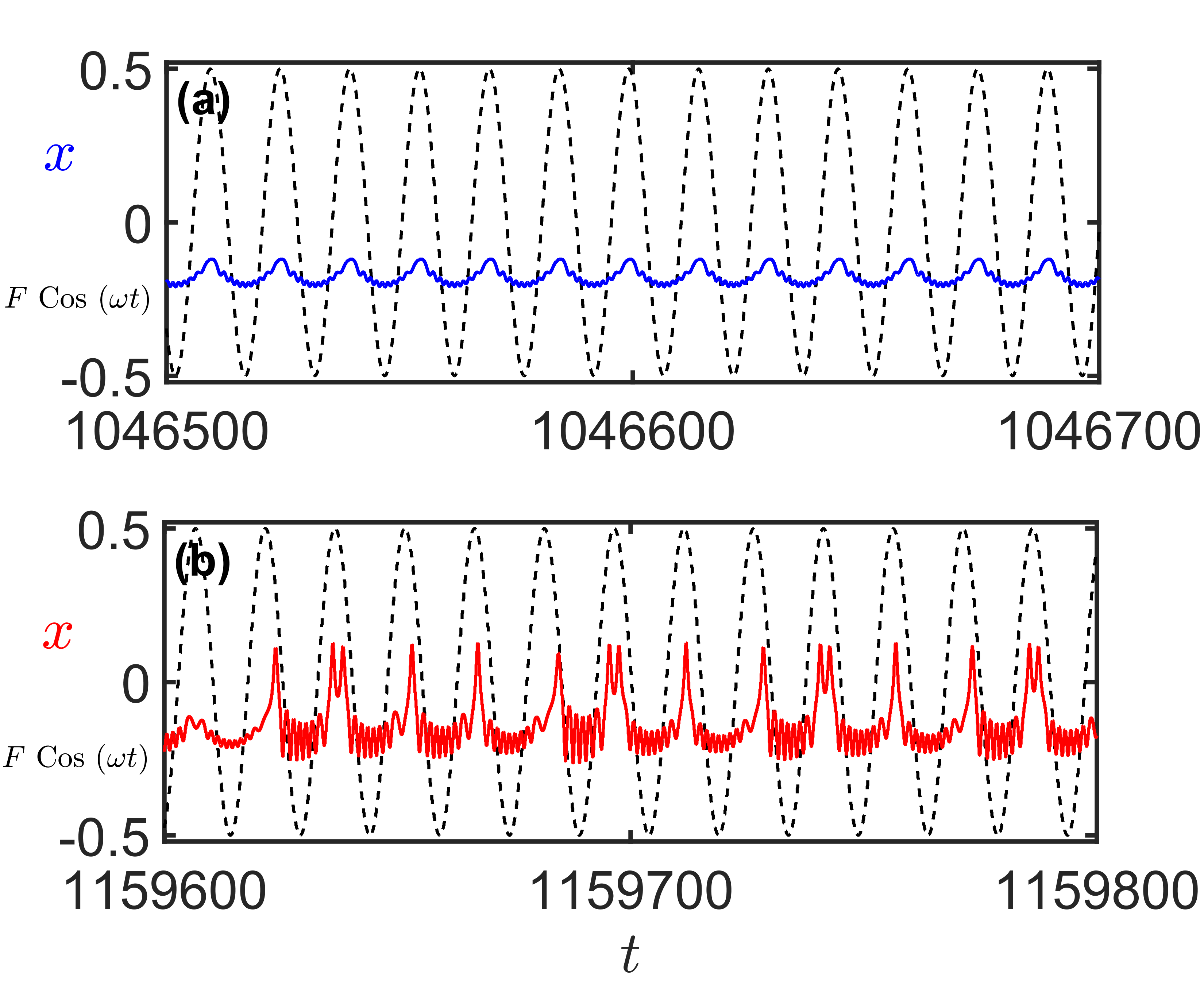}
\caption{Time series of the system (\ref{mastereqn}) at $c_2=5.6494884$ superimposed over the time series of the external drive for (a) non-extreme regions and (b) extreme regions. Blue solid curves represent the time series of the system during periodic MMO exhibition, Red solid curves represent the time series of the system during extreme MMO dynamics, and the black dotted curves represent the time series of the external driving force.}
\label{force-mecha}
\end{figure}

 Next, to investigate the effect of the external drive on the dynamics, we superimpose the time series of the driving force over the time series of the system response (i.e. the solution of  Eq.~(\ref{mastereqn})). The plots for the two distinct cases corresponding to the absence and emergence of extreme events are presented separately in Fig.~\ref{force-mecha}(a) and Fig.~\ref{force-mecha}(b), respectively. In the case where extreme events are absent, we observe from Fig.~\ref{force-mecha}(a) that as the external drive rises to its maxima (dotted black curves), single large amplitude oscillations occur, whereas, for decreasing values of the external drive towards minima, eight small amplitude oscillations occur (solid blue curves). Note that all these occur for dynamics within the same (left) well. However, in the other case spanning extreme irregular MMO bursts, we find that the increase and decrease in the forcing amplitude drives the system intermittently back and forth between the wells, marking the emergence of extreme events. In other words, during the burst of extreme MMO, whenever the amplitude of the driving force has a positive slope (i.e. increasing amplitude in the dotted black curves in Fig.~\ref{force-mecha}(b)), the trajectory of the system (solid red curves in Fig.~\ref{force-mecha}(b)) moves into the right well exhibiting large amplitude peaks, and returns to the left well during the descending phase of the driving force (negative slope). This feature is depicted in Fig.~\ref{force-mecha}(b), wherein the red solid curves represent the system dynamics during the extreme MMO, and the black dotted curves portray the external driving force.

\par In light of the results presented, we can now understand the dynamic mechanism that explains the emergence of extreme events observed in our results. At the outset, we note that for zero forcing and for the chosen values of $c_1$, $c_2$, and $c_3$, there occurs a double well in the potential with a single saddle point at $(0,0)$ in the phase space along with two stable foci at $(0.01030,0)$ and $(-0.18156,0)$. Moreover, we know that the value of $c_2$ determines the symmetry and depth of the left potential well (for negative $c_1$). Under excitation by an external drive and increasing $c_2$, we observe from Fig.~\ref{figbifur} that the dynamics transition from chaotic to exhibiting multiperiodic oscillations at $c_2=91.5494885$. For the parameter values chosen, the dynamics are mostly confined to the left well, whilst intermittent inter-well hopping occurs chaotically. However, as $c_2$ increases, the time spent by the trajectory in the right well during the inter-well excursions gradually reduces. This occurs due to the increase in the depth of the left potential well. More specifically, a deeper left potential well implies that the system requires more energy to cross the unstable fixed point at $(0,0)$ and successfully transition to the right potential well. The reduction in time (spent in the right potential well) occurs monotonically until $c_2=91.5464884$, after which, at $c_2=91.5464885$, the system is entirely unable to cross the unstable fixed point at $(0,0)$. Therefore, the system is confined to the left potential well. Moreover, we find that these transitions are rare enough to authentically qualify as rare events as per the POT criterion. 
\par
\begin{figure}[!h]
\includegraphics[width=1.0\linewidth]{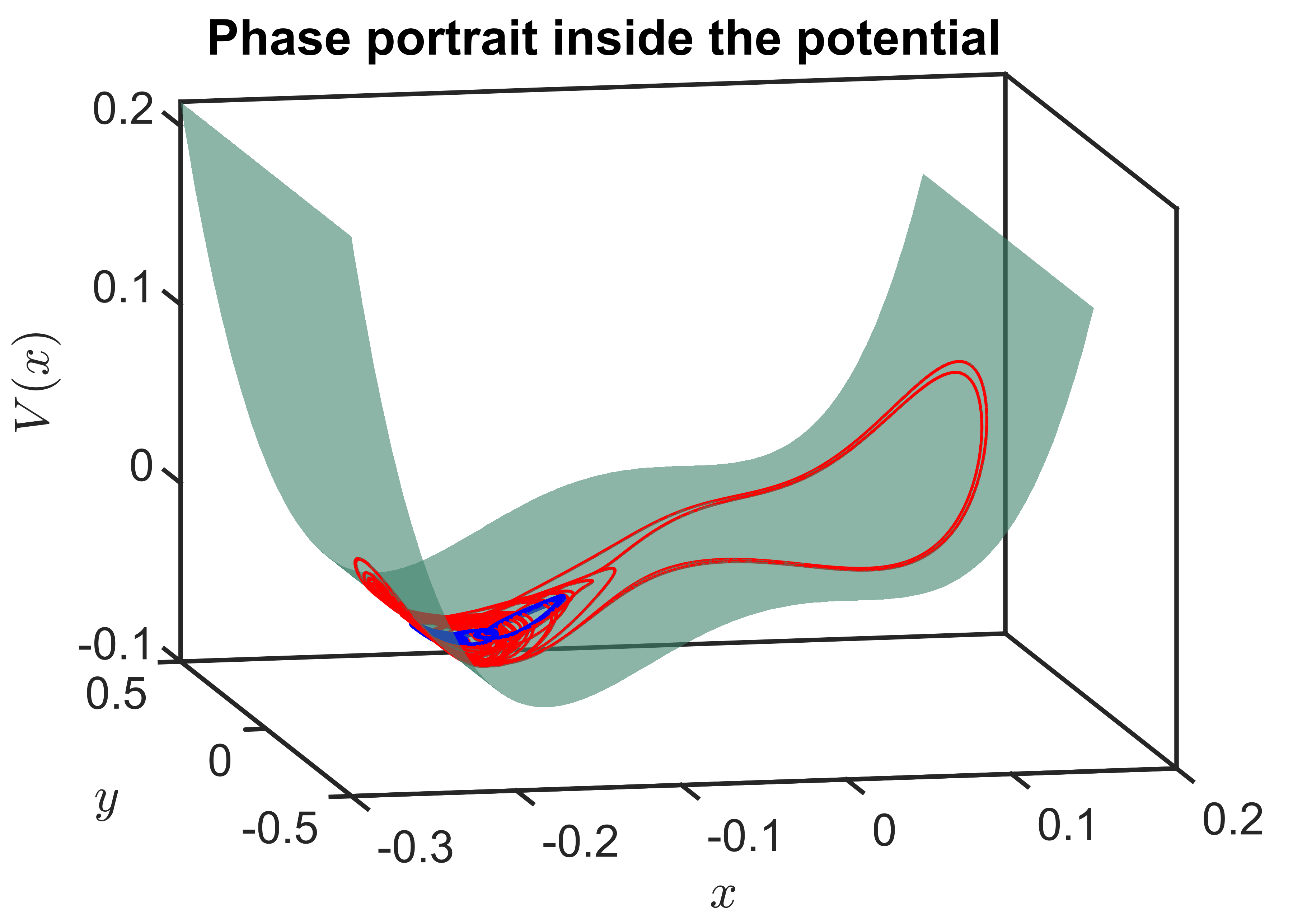}
\caption{Depiction of the Phase portrait of system (\ref{mastereqn}), inside the asymmetric double-well potential $V(x)$. Blue trajectories represent bounded motion inside the left well, and red trajectories represent excursions to the right well for short periods of time.}
\label{poten_port}
\end{figure}

 To conclude the Section with a qualitative summary, extreme irregular MMO bursting emerges in the H-D oscillator due to rare, chaotic inter-well hopping of the trajectories; the hopping essentially involves crossing the unstable fixed point at the origin in phase space. Furthermore, we would like to highlight yet another feature of the dynamics that we find remarkable. In a system with a double well potential, hopping of trajectories between the wells typically results in chaotic oscillations for the entire trajectory, i.e. for the entire path encompassing segments in both the left and the right wells. However, in the present case, the system exhibits periodic MMO when the trajectories remain inside the left potential, while the response becomes chaotic when the trajectory hops back and forth between the two wells. Specifically, such response manifests as extreme bursting MMO events. To clarify the mechanism further, we overlay the phase 
 portrait on the potential ($V(x)$ presented in Sec.~\ref{model}) in Fig.~\ref{poten_port}. The blue curves represent the regular MMO bounded inside the left well. The red curves represent the irregular MMO hopping back and forth between the wells. Moreover, the red curves also highlight the rarity of such hopping.

\section{Reliable indicator: Velocity}
\label{ri_sec}
An important question that follows the identification of extreme events in a dynamical system is that of predicting their emergence. While it is typically arduous to identify reliable predictors, forecasting extreme events is important, especially in cases where interventions guided by predictions can help mitigate their adverse consequences. Key prognosticators in this context - in cases where they are measurable and reliable - are analytical measures computed from system response termed \textit{reliable indicators}. In the present case of an extreme MMO burst, fortunately, we find that the velocity of the Helmholtz-Duffing oscillator can be used as a reliable measure to track and predict extreme events in advance, as detailed below.

\begin{figure}[!h]
\includegraphics[width=1.0\linewidth]{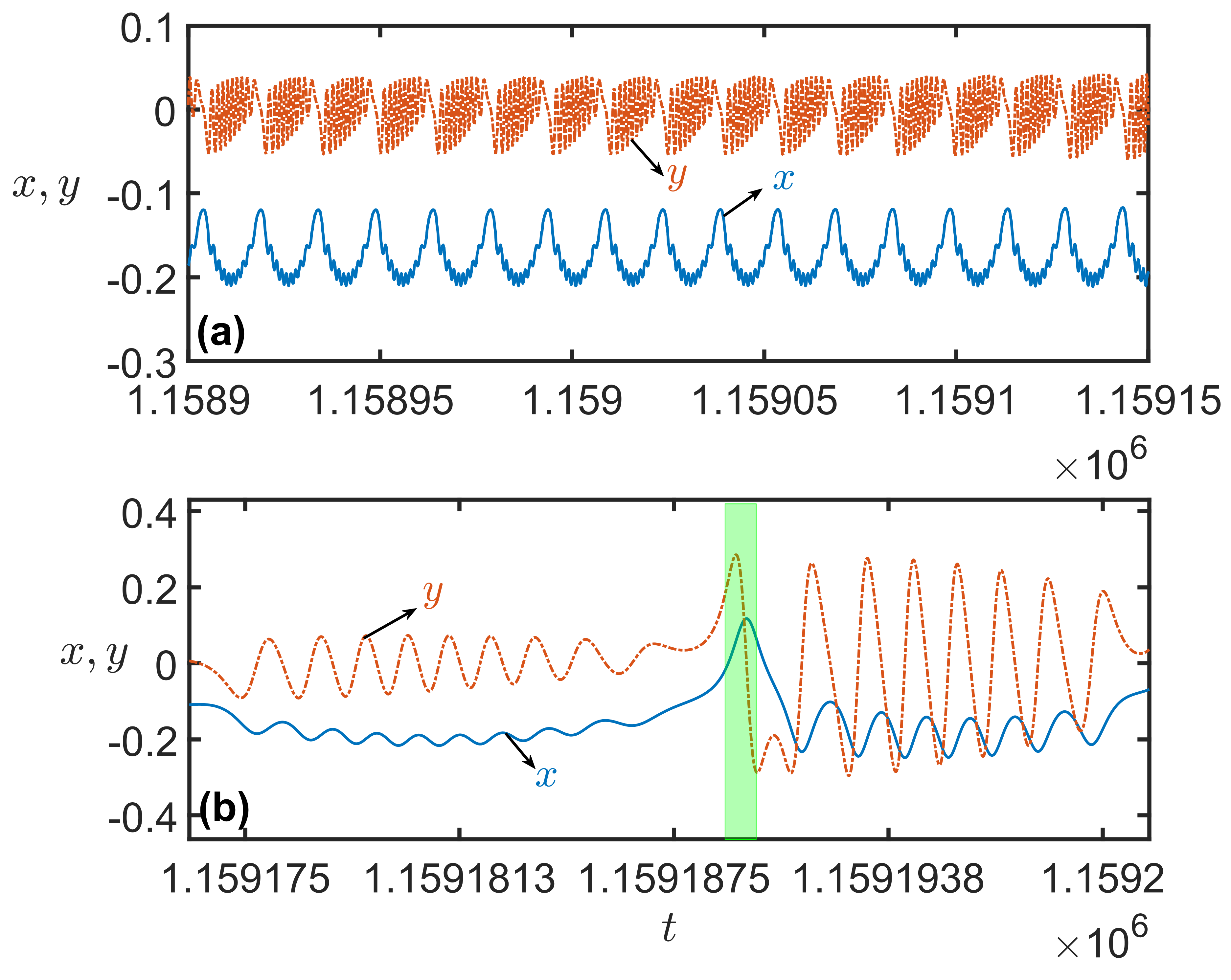}
\caption{Time series of $x$ (blue curve) and $y$ (orange curve) (a) before the emergence of extreme bursts and (b) during the emergence of extreme bursts depicting the shooting of the $y$ variable preceding the $x$ variable. Both the plots represent the dynamics of the system (\ref{mastereqn}) at $c_2=91.5464884$.}
\label{ri}
\end{figure}

We observe from the results presented in Fig.~\ref{ri}(a) that as the $x-$variable (position) of the  H-D oscillator exhibits periodic MMO, concurrently, the $y-$ variable (velocity) exhibits periodic bursting oscillations. Note that the dash-dotted orange curve is the time series of the $y-$variable while the solid blue curve is the time series of $x-$variable. There exists a constant spatial difference between the trajectories of both variables, particularly in Fig.~\ref{ri}(a), where both variables are periodic. However, interestingly one observes from the results in Fig.~\ref{ri}(b) that shortly preceding the emergence of the extreme MMO in the $x-$variable, the $y-$variable shoots up to a higher value, deviating from its periodic burst. Imminently, an extreme MMO burst occurs in the $x-$variable. Moreover, as soon as the $y-$variable returns to the periodic bursting, the $x-$variable also returns to its periodic MMO. Therefore, every departure of the $y-$ variable from its periodic burst to a higher value is suggestive of an imminent extreme MMO burst in the $ x-$ variable. Furthermore, we present this more precisely in the green shaded region in Fig.~\ref{ri}(b), where the increase in the $y$ variable clearly precedes the increase in the $x-$ variable. We find this significant feature to be consistently true with every occurrence of extreme MMO bursts in our results and, therefore, conclude that the velocity is a reliable lead indicator of extreme MMO bursts in the H-D oscillator.

\section{Discussion and Conclusions}
\label{discuss}
In nonlinear dynamical systems evolving under a double-well potential, transitions of the system trajectory between the wells is a well-recognized phenomenon, both in the cases of deterministic and stochastic motion. In deterministic scenarios, such dynamics are typically chaotic, also often under the influence of external excitation. Moreover, when the double-well potential is asymmetric, rare inter-well transitions that coexist along with chaotic motion in the predominantly occupied well have also been reported. In the present case of the H-D oscillator dynamics, our analysis uncovered multiperiodic dynamics for higher values of the quadratic nonlinear coefficient that characterizes the depth of one of the potential wells. Moreover, we reason that, while lower values of the quadratic nonlinear coefficient allow inter-well transitions (across the barrier of the unstable fixed point), higher values of the former impose correspondingly higher energy thresholds that challenge such transitions. We underscore that the striking feature of the inter-well hopping reported here is the exhibition of a periodic MMO in the deeper asymmetric well before the hopping. Intermittently, the trajectories hop back and forth while exhibiting irregular MMO, before returning to the deeper well to exhibit periodic MMO. Indeed this intermittent, irregular MMO burst manifests as rare extreme events. Also, a further decrease in the value of the quadratic nonlinear coefficient increases the duration of the irregular MMO hop, thereby foreclosing the possibility of extreme events. Furthermore, we interestingly find that such a rare irregular MMO burst consistently follows a sudden upshot in the system velocity. This substantiates the important conclusion that the velocity of the H-D oscillator can be used as a reliable lead indicator for predicting the emergence of the novel type of extreme event that we identify in the H-D oscillator.

In this article, we present a new class of extreme events which we propose to term \textit{extreme irregular mixed mode oscillatory bursts}. We report the latter emerging in a driven H-D oscillator with an asymmetric double-well potential. The asymmetry in the depths of the potential wells, along with the external drive, play vital roles in engendering these extreme bursts of irregular MMO. Extreme bursts are classified using the POT method, after which they are also found to fit the GEV distribution. Moreover, the dynamical mechanism underlying the emergence of the identified extreme events is determined using appropriate analysis. Furthermore, the system velocity is identified as a reliable lead indicator to predict extreme events in the system. We conclude with the hope that the \textit{extreme irregular MMO burst} reported in this article will open new pathways in the theoretical analysis of extreme events in nonlinear dynamical systems and will also introduce new perspectives in the characterization and control of this important dynamic phenomenon in multiple applications, including in those where MMOs are an important consideration. 
\vspace{0.1cm}
\section*{Acknowledgments}
S.S. thanks the Science and Engineering Research Board (SERB), Department of Science and Technology (DST), Government of India for financial support in the form of a National Post-Doctoral Fellowship (File No.~PDF/2022/001760). S.R. acknowledges funding from the US National Science Foundation (Award No: CMMI 2140405). 

\bibliography{ref.bib}
\end{document}